\documentclass[runningheads]{llncs}
\usepackage{graphicx}
\usepackage[export]{adjustbox}
\usepackage{xcolor}
\usepackage{enumitem}
\newenvironment{longdescription}
  {\begin{description}[style=unboxed]}
  {\end{description}}
\usepackage{hyperref} % added to avoid problems with symbols in URLs
\usepackage{subfig}

\begin{document}
\title{Water Quality of Lake Mjøsa through Satellite Images: A Preliminary Study}
\author{Anders Gjesdal Oliversen \and
Hilda Deborah}
\authorrunning{A. G. Oliversen and H. Deborah}
\institute{Department of Computer Science, Norwegian University of Science and Technology\\
Teknologiveien 22, 2815 Gjøvik, Norway\\
\email{andergol@stud.ntnu.no, hilda.deborah@ntnu.no}\\
\url{https://www.ntnu.edu/colourlab}}
\maketitle              % typeset the header of the contribution
\begin{abstract}

Since around 2010, the water quality in lake Mjøsa has begun to decline after an earlier successful effort to improve the water quality (\textit{Mjøsaksjonen}).
In this study, we investigate the possibility of using satellite imagery to monitor the water quality of lake Mjøsa, and also compare the use of such an approach to current methods for monitoring the lake. Using satellite images for remote sensing has layers of complexity that were not apparent to us at first, and this paper summarizes some of our findings by discussing factors of water quality, state of the art for use of satellite imagery for water quality and some of our own initial results.
While our results where not too accurate for this preliminary study, it gives an introduction to the field and some of the methods that may bring better results for further research. Satellite images can play an important role in monitoring changing waters. The field is in development and many factors of water quality may be analysed with increasing levels of accuracy.

\end{abstract}

%%%%%%%%%%%%%%%%%%%%%%%%%%%%%%%%%%%%%%%%%%%%%%%%%%%%%%%%%
\section{Introduction}
%%%%%%%%%%%%%%%%%%%%%%%%%%%%%%%%%%%%%%%%%%%%%%%%%%%%%%%%%
About 100 000 people have Mjøsa as their source of drinking water \cite{FaktaMjosa}, and people rely on Mjøsa for recreational purposes like swimming, sailing and fishing. Many animal species are also dependent on Mjøsa, and as the largest lake in Norway it plays an important role in the ecosystem of the area.

Historically, the water quality of Mjøsa has not always been good. One of the first records to human pollution of Lake Mjøsa affecting the water quality is from 1872~\cite{Nashoug1999}, when sewage water released directly into Mjøsa without any form of prior cleansing treatment led to typus epidemic in Hamar that lasted 10 years. Gjøvik was also struck by a similar typhus epidemic in 1931, resulting from hospital sewage released into Mjøsa~\cite{Nashoug1999}.
From the mid 1950s and into the 1960s, toxic algae blooms came with increasing frequency and systematic monitoring of the water quality of lake Mjøsa and the tributary rivers commenced. The Mjøsa Campaign, or \textit{Mjøsaksjonen}, was initiated in 1973 as an attempt to combat the pollution and \textit{save} lake Mjøsa. It was a successful effort and phosphorus released into Mjøsa per year was reduced from 459 tons in 1972 to 225 tons in 1980~\cite{Nashoug1999}. 
The condition of Mjøsa has been evaluated as “acceptable” in most of the years after around 1990, but this started to change again around 2010. Several large algae blooms have been reported in recent years \cite{NRK2014,NRK2019,NRK2021}, and levels of phosphorus found in the water have been increasing \cite{Thrane2021}. A massive cyanobacteria bloom has also been reported in 2019 \cite{Thrane2021}.

Traditionally, water quality analysis has been done through field sampling and laboratory analysis. This approach has its limitation especially for the monitoring large areas and numbers of water bodies, because the distribution of nutrients, algal blooms, and suspended matter are patchy \cite{Song2011}. In this preliminary work, we are investigating the possibilities of using freely available Sentinel-2 images to monitor the water quality in Mjøsa. We aim to give an introduction to the topic, as well as report the preliminary results, challenges, and lessons learnt in this study. This article will be especially useful for readers from data science or computer science, who do not have prior domain knowledge to work in the topic of remote sensing for water quality.

%%%%%%%%%%%%%%%%%%%%%%%%%%%%%%%%%%%%%%%%%%%%%%%%%%%%%%%%%
\section{Environmental Water Quality}\label{sec:waterq}
%%%%%%%%%%%%%%%%%%%%%%%%%%%%%%%%%%%%%%%%%%%%%%%%%%%%%%%%%
\textit{Water quality} is defined as a measure of the condition of water relative to the requirements of one or more species and/or to any human need or purpose \cite{Johnson1997}. The U.S. Environmental Protection Agency also provides a definition for \textit{water quality criteria} as the levels of water quality expected to render a body of water suitable for its designated use \cite{USEPA1994}. It was further elaborated that the criteria are based on specific levels of pollutants that would make the water harmful for a variety of purposes. From both definitions, we know that if we are to assess water quality, we need to determine and specify the purpose. In this work, we are specifically working with environmental water quality and, therefore, are concerned with evaluating \textit{environmental water quality indicators}. From this point, we will use the term water quality to specifically the environmental one, and not relating it to concerns of consumption or domestic/ industrial uses. 

%%%%%%%%%%%%%%%%%%%%%%%%%%%%%%%%%%%%%%%%%%%%%%%%%%%%%%%%%
\subsection{Eutrophication and Trophic State Index (TSI)}
Nutrient pollution in water, i.e., when too many nutrients are added to the body of water, will cause excessive growth of algae. Algae and plants are dependent on nitrogen and phosphorus for growth, and these nutrients are usually scarce in natural environments. High amounts of nitrogen and phosphorus can make algae and water plants grow faster than the ecosystem can handle~\cite{NOAA_eutrophication,Schindler2006}. Therefore, it is important to monitor \textit{eutrophication}. The word eutrophication comes from the Greek \textit{eutrophos}, meaning “well-nourished”~\cite{MW_eutrophication}. When a body of water is enriched with nutrients and minerals it is described as the eutrophication of that body of water. Phosphorus has long been seen as the main nutrient controlling freshwater productivity, while controlling nitrogen has been seen as more important for coastal waters. Some recent papers argue that control of both nutrients is required to manage eutrophication in both freshwater and coastal waters~\cite{Paerl2009}. 

Trophic means something of or relating to nutrition~\cite{MW_trophic}, and the Trophic State Index (TSI) is a system used to rate water bodies based on their nutritional state~\cite{LCWA_TSI}. Water bodies with low levels of nutrients will have low amounts of biological activity, while those with high levels of nutrients will have high amounts of biological activity. 
TSI rates water bodies as \textit{oligotrophic}, \textit{mesotrophic}, \textit{eutrophic} or \textit{hypereutrophic} depending on the amount of dissolved nutrients in the water. Where oligotrophic water bodies have the least amount of biological activity, and hypereutrophic have the most. 
The low biological activity in oligotrophic lakes means they have low algal production, clear waters with good visibility and high drinking water quality.

%%%%%%%%%%%%%%%%%%%%%%%%%%%%%%%%%%%%%%%%%%%%%%%%%%%%%%%%%
\subsection{On case 1 and case 2 waters}
A distinction is made between case 1 and case 2 waters because they have different optical properties and, consequently, different algorithms are required to work with them. {Case 1 waters} are typically {open oceans} or {oligotrophic lakes}. {Case 2 waters} are more complex and are typically {coastal waters} and {lakes}. However, while still in use, the case 1 and case 2 separation may be too much of a simplification and it might be better if the optical properties of lakes are determined on an individual basis~\cite{Mobley2004}.

%%%%%%%%%%%%%%%%%%%%%%%%%%%%%%%%%%%%%%%%%%%%%%%%%%%%%%%%%
\subsection{Water quality indicators}\label{subsec:indicators}
Indicators for determining water quality are many and can be grouped into biological, chemical, and physical indicators. However, only a part of those indicators will be relevant for optical sensors. The following are bio-optical indicators or properties that can be monitored by means of remote sensing:

\begin{longdescription}
    \item[Chlorophyll-a (chl-a)]is one of the most common indicators of water quality looked for with remote sensing. It is the primary indicator of phytoplankton production, and phytoplankton blooms caused by eutrophication can be problematic. Chl-a absorbs light strongly at the blue (near 440 nm) and the red (near 670 nm) parts of the light spectrum~\cite{Mishra2017}. All green plants and algae contain chl-a. So while analysing chl-a content in a body of water can be useful, it is not very specific.
    
    \item[Phycocyanin] is a pigment unique to cyanobacteria and has distinctive optical features around 620-630 nm~\cite{Mishra2017}. Cyanobacteria ("blue-green algae") toxins can cause severe health issues~\cite{USEPACyano,Hilborn2014}. Cyanobacteria also contain chl-a, but looking for chl-a is not a good way to identify cyanobacteria since all phytoplankton contains chl-a. Cyanobacteria blooms have been problematic in Mjøsa, and it would be useful to be able to distinguish between cyanobacteria and other algae.
    
    \item[Total Suspended Solids (TSS), Colored Dissolved Organic Matter (CDOM), and turbidity] all have an impact on reflectance. This makes them feasible to measure with remote sensing, but if the reflectance changes the atmospheric correction and algorithms for analysing the water may also need to be changed~\cite{Mishra2017}.
    TSS, CDOM and turbidity all impact the clarity of water in different ways, and different methods would be used to measure each of them. Clear water is generally an indicator of healthy water, and a sudden decrease in clarity in a previously clear body of water (like Mjøsa) may indicate that something is wrong~\cite{Fondriest2014}.
    
    \item[Aquatic plants.] Changing plant biomass can impact a lake’s biological communities, water, sediment chemistry, and nutrient cycling. Eutrophication can reduce the amount of (submerged) aquatic plants due to reduced light penetration. Eutrophication can also stimulate aquatic plant growth until they become a nuisance present in excessive amounts.
    A number of indices have been developed to detect aquatic plants, but it was concluded that the work done so far is incompatible and inconsistent~\cite{Mishra2017}. The problem is not to detect if there are plants in the water, but trying to differentiate between them and identify problematic plant species. Some plants may be native to a body of water and are no cause for concern, while invasive species of water plants can be very problematic.  
\end{longdescription}

%%%%%%%%%%%%%%%%%%%%%%%%%%%%%%%%%%%%%%%%%%%%%%%%%%%%%%%%%
\subsection{Water quality in Mjøsa in recent years}

The Mjøsa Campaign initiated in 1973 focused on reducing phosphorus in Mjøsa, because the access to phosphorus in Mjøsa is the limiting growth factor for algae in the lake. Recently, however, not only the levels of phosphorus that have been increasing but also the levels of nitrogen in the water~\cite{Thrane2021}. Nitrogen from Mjøsa will eventually end up in Oslofjorden via Glomma, and can potentially have a larger negative impact on water quality in the coastal waters there~\cite{Nashoug1999}.

A recently published report~\cite{Bechmann2021} found the main sources contributing phosphorus to lake Mjøsa to be agriculture and sewage. Agriculture generally contributes phosphorus with low bioavailability, while sewage contributes phosphorus with high bioavailability. Bioavailable phosphorus is more easily transported in the water and is more accessible for algae growth. Sewage is estimated to contribute 50-75\% of the total amount of bioavailable phosphorus in the rivers running into Mjøsa. 
Concentration of E. coli have also been found to be very high in many of the tributary rivers to Mjøsa in recent years~\cite{Bechmann2021}. This indicates that the water quality in the rivers is affected by sewage and livestock manure. E. coli concentrations were found to be above the set limit for use in soil irrigation in many of these rivers~\cite{Bechmann2021}. 

From 1972 to 2020 the mean temperature in the upper water layer of Mjøsa for the period June-October has increased by 2.0 °C, and the maximum temperature has increased by 3.3 °C~\cite{Thrane2021}. Higher temperature in the lake could lead to more eutrophic conditions. There has also been to be more flooding in the region in recent years and this leads to higher total loads of phosphorus in Mjøsa and the tributary lakes. Heavy rain followed by warm weather results in favourable conditions for algal growth, and more of this kind of weather is expected in the region in the future due to climate change. This will make it extra challenging to reduce the pollution of lake Mjøsa in the coming years.

%%%%%%%%%%%%%%%%%%%%%%%%%%%%%%%%%%%%%%%%%%%%%%%%%%%%%%%%%
\section{Materials and Methods}
%%%%%%%%%%%%%%%%%%%%%%%%%%%%%%%%%%%%%%%%%%%%%%%%%%%%%%%%%
In this section we provide details regarding data and methods used in this study, where we focus on lake Mjøsa. Relating to the different aspects that have been described in Section \ref{sec:waterq}, the characteristics of Mjøsa are as follows. Mjøsa’s natural state is that of an oligotrophic lake~\cite{Nashoug1999}. However, regarding the type of waters, Mjøsa is probably closer to the case 2 type of waters due to the ongoing pollution.

%%%%%%%%%%%%%%%%%%%%%%%%%%%%%%%%%%%%%%%%%%%%%%%%%%%%%%%%%
\subsection{Data}

\subsubsection{Sentinel-2 images}
Sentinel-2 is part of the Copernicus programme by The European Space Agency (ESA) and environmental data from the programme has free, full and open access to all users, and we use the data for this preliminary study. Data from other satellites could have been used as well, but working with different sensors will further complicate the work and was decided against for this preliminary study.
Sentinel-2 is a constellation of two identical satellites in the same orbit developed and operated by the European Space Agency. Sentinel-2A launched 23rd June 2015 and Sentinel-2B launched 7th March 2017. The revisit time every 1-5 days, but more frequent at high altitudes like Norway and there is Sentinel-2 data of lake Mjøsa every 1-2 days. The satellites have 13 spectral bands with spatial resolutions of 10m, 20m and 60m.

\subsubsection{In-situ measurement}
\begin{figure}[!ht]
\centering
\includegraphics[width=.6\textwidth]{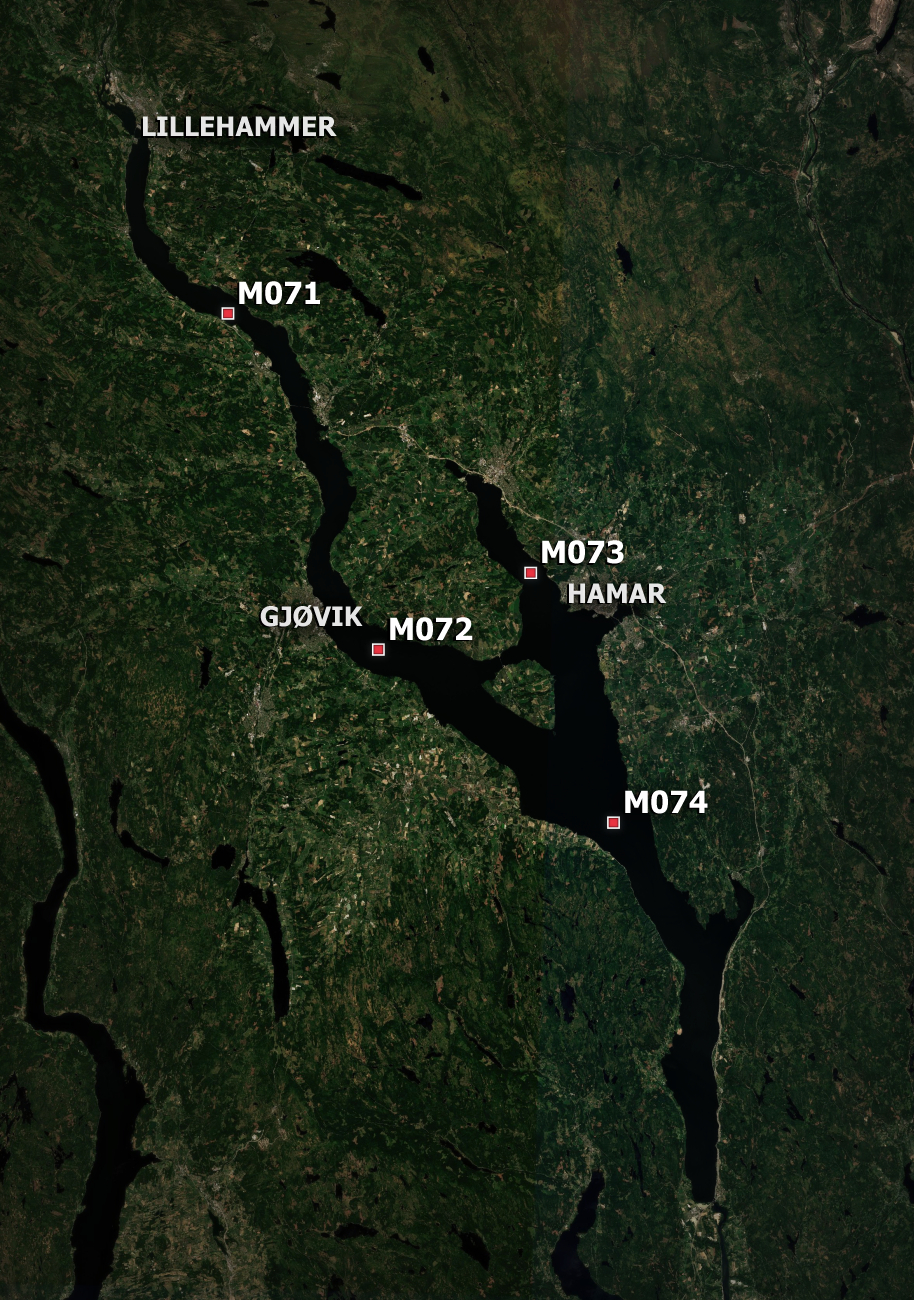}
\caption{Satellite image of Mjøsa with NIVA stations for monitoring water quality.} 
\label{figNivaSta}
\end{figure}

NIVA has been monitoring the water quality of Mjøsa since 1972. Physical data gathering is done on location by NIVA researchers using various equipment and samples are brought to laboratories for analysis. Monthly tests are done in the months of May to October.
NIVA shared some of the data gathered in 2016-2020 with us for comparison with satellite data. The data was on turbidity, chlorophyll-a and total organic carbon (TOC) gathered from four stations in Mjøsa, see Fig.~\ref{figNivaSta}.

%%%%%%%%%%%%%%%%%%%%%%%%%%%%%%%%%%%%%%%%%%%%%%%%%%%%%%%%%
\subsection{Method}

\subsubsection{EO Browser}
is a cloud platform allowing the user to easily access high-resolution data from many different satellites with almost no loading time, and it is possible to do simple scripting directly in EO Browser.
Analysing the data directly in EO browser would save a lot of time and storage space compared to downloading data from sources like Copernicus and using other tools for analysis. The data is big, and if analysis is to be done on a regular basis the time and space saved with a tool like EO Browser would be significant. 
Although EO Browser can be a great tool, there were a few challenges with using it for analysing water quality that we found during this study:
\begin{description}
  \item[Image stitching] EO Browser stitches multiple images together. This effect is hardly visible in true colour images, but creates problems with overlapping areas when processing the data. There is a stitch that often happens in the middle of Mjøsa, making data in that area useless for analysis.
  \item[Limited scripting tool] EO Browser has good scripting capabilities for manipulating the different spectral bands, but usually simple tasks like extracting numerical values can be difficult.
  \item[Sen2Cor Atmospheric Correction] Only Sentinel-2 data with the Sen2Cor atmospheric correction is available directly in EO Browser. This will be fine in most cases if analysing land areas, but it is not regarded as a good atmospheric correction to use for analysing water quality. This will be explained in more details later in this section.
\end{description}

\subsubsection{Water quality scripts}
There are several Sentinel-2 scripts available for water analysis in the Sentinel Hub custom scripts repository, that can be used on EO Browser. Below are the ones we used.
\begin{longdescription}
  \item[Se2WaQ - Sentinel-2 Water Quality Script] \cite{Toming2016,Potes2018} was one of the scripts we found to be promising in the initial stages of the project and decided to use for comparison with in situ data. Se2WaQ can displays the spatial distribution of six indicators of water quality, i.e., the concentration of Chlorophyll-a (chl-a), the density of cyanobacteria, turbidity, colored dissolved organic matter (CDOM), Dissolved organic carbon (DOC), and Water color. Of these indicators, however, we only had in situ data for chl-a and turbidity to compare with.
  
  \item[APA - Aquatic Plants and Algae Custom Script Detector] \cite{APAScript} detects aquatic plants and algae in water bodies. We did not have in situ data from NIVA to directly compare with regarding these factors. Thus, we compared the satellite data with periods of algal blooms reported in the news.
\end{longdescription}

%%%%%%%%%%%%%%%%%%%%%%%%%%%%%%%%%%%%%%%%%%%%%%%%%%%%%%%%%
\subsection{Challenges in the remote sensing of inland waters}
Previously, we have shortly touched upon the challenges we encountered while using the EO Browser. There are, however, important considerations that are specific to the remote sensing of inland waters. Below we elaborate in more details the two main challenges.

\subsubsection{Atmospheric correction} is an essential step before analysing satellite data. We have to remove the influence of atmosphere and surface to obtain the reflectance of interest, i.e., of the water itself.
One of the atmospheric phenomena that complicates the atmospheric correction for inland and coastal waters is the \textit{adjacency effect}, i.e., where scattering in the atmosphere contaminates water pixels with radiation from the surrounding land~\cite{Mishra2017}. It has also been shown that all atmospheric correction algorithms for Sentinel-2A have high uncertainties~\cite{Warren2019}. In the evaluation, Polymer and C2RCC achieved the lowest mean square differences.
Sen2Cor, which is the default atmospheric correction for Sentinel-2 and the one we used, was said to have generally poor performance for water bodies, but better performance for inland waters than for coastal waters. 

\subsubsection{Cloud detection} 
\begin{figure}[t]
    \centering
    \subfloat[Sentinel-2 image]{\includegraphics[width=.7\linewidth]{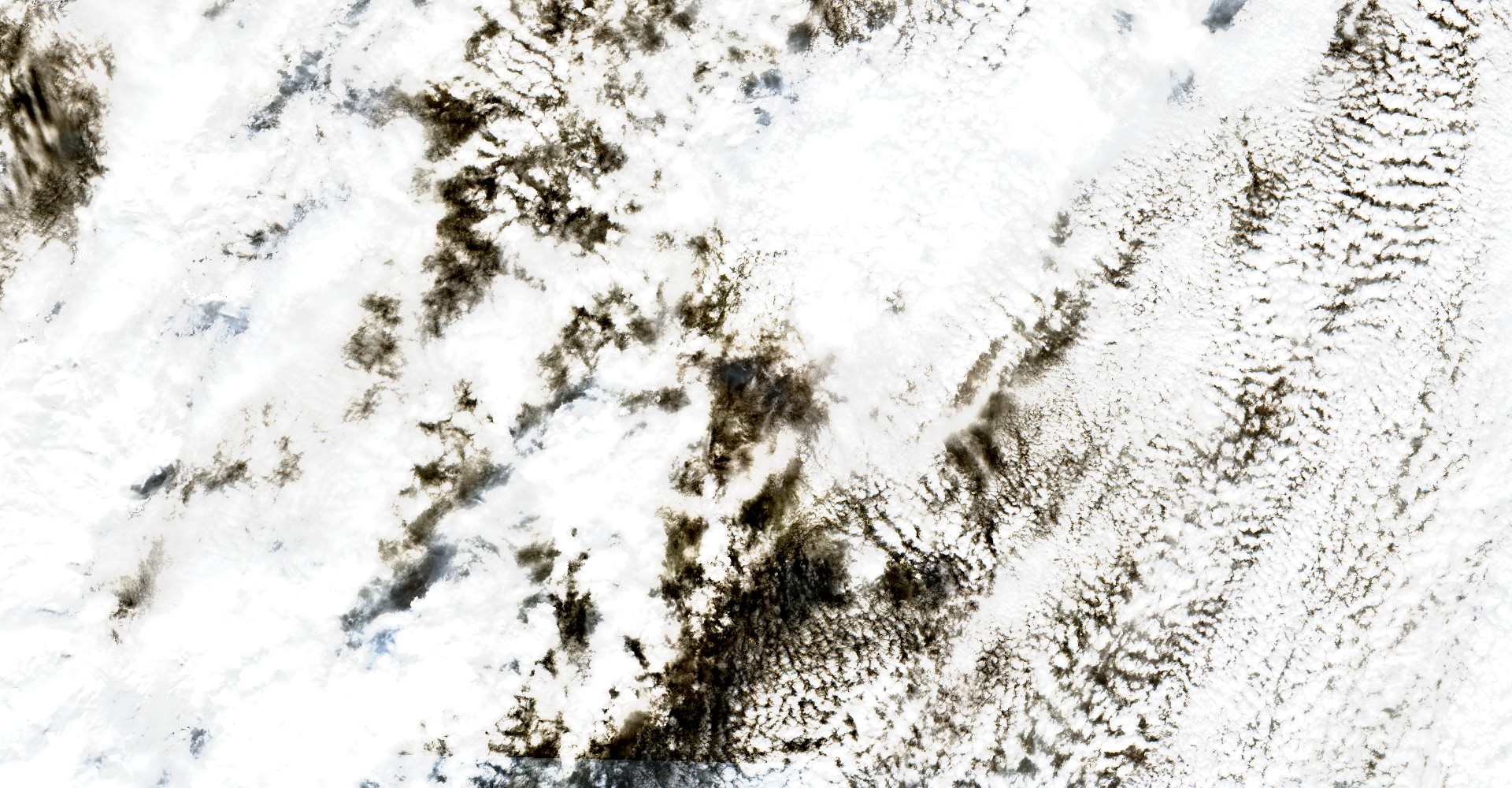}}\\
    \subfloat[Radar image]{\includegraphics[width=.7\linewidth]{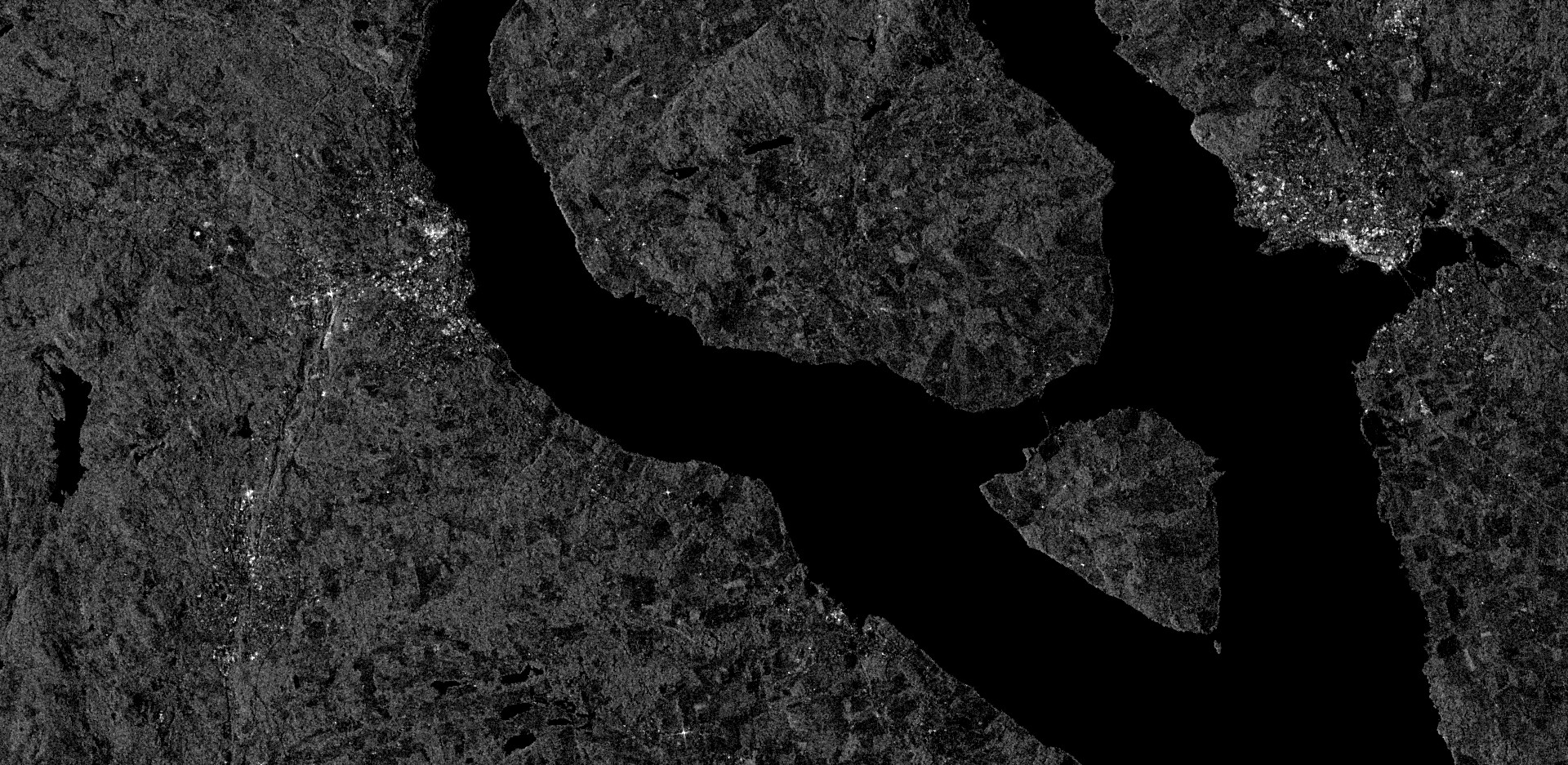}}
    \caption{Images taken over Mjøsa on 30th September 2019. Clouds completely cover the Sentinel-2 image but the radar image of the same location on the same date completely sees through the clouds.} \label{figRadar}
\end{figure}

is also essential since clouds often fully or partially obscure the data when using satellite images. Clouds obscuring the entire image is a common occurrence with satellite data and this is one of the greatest challenges when working with satellite images. If clouds only partially obscure the data, they only need to be detected and masked out so they do not interfere with the usable data. Cloud shadows can also distort the data and should be masked out. There is radar satellite imagery available and this can be a solution for certain uses. Radar can see through clouds, see Fig.~\ref{figRadar}, and for example detect larger floods or ice cover. Unfortunately {radar cannot be used to analyse parameters of water quality}. So this is ruled out as a potential solution in our current study. 

\begin{figure}[t]
    \centering
    \subfloat[Original cloudy image]{\includegraphics[width=.47\linewidth]{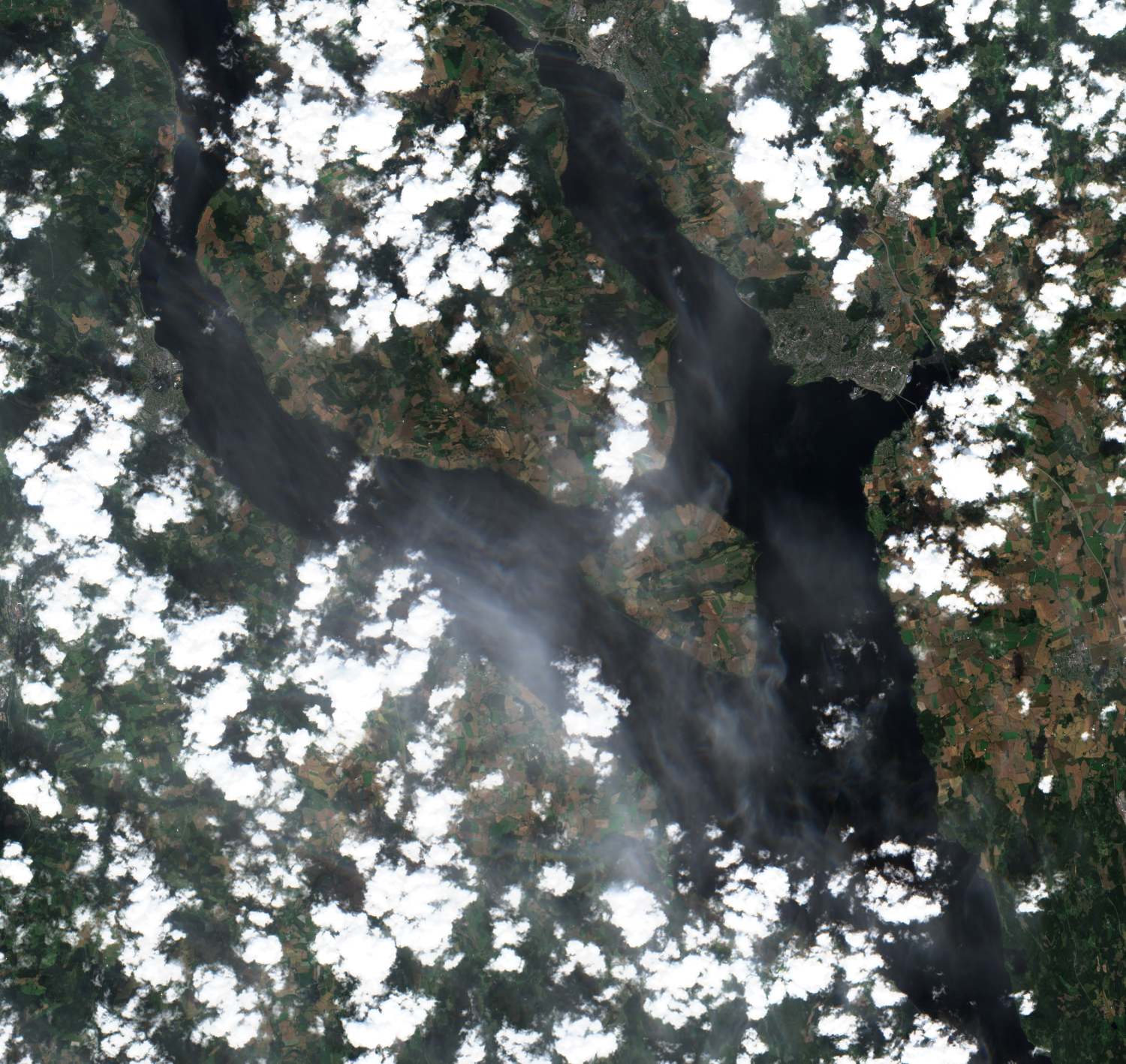}}\hfil
    \subfloat[Cloud detected image]{\includegraphics[width=.47\linewidth]{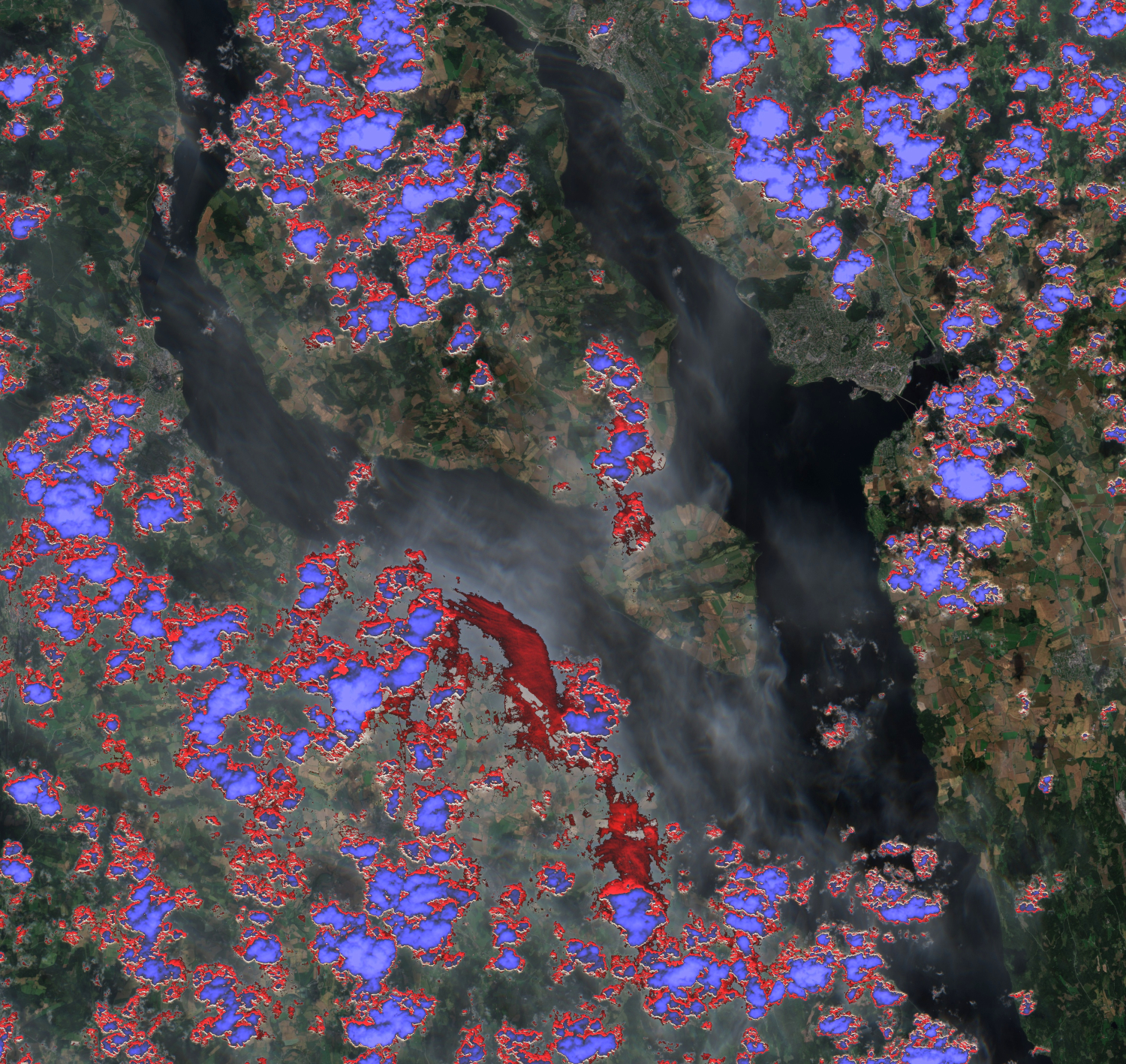}}
    \caption{Result of cloud detection from a satellite image using Braaten-Cohen-Yang detector. Thick cloud is segmented as purple pixels, while the thin ones with red colors.}
    \label{figCD}
\end{figure}

\noindent There is a fine balance in cloud detection between {failing to remove contaminated data and removing usable data}. A recent study compared cloud detection algorithms for Sentinel-2 and found that the LaSRC algorithm was the best one for finding clouds and Tmask algorithm was the best for finding shadows \cite{Tarrio2020}. Improvements in clouds detection are being made, but for now it is probably best if a human looks over to check if the cloud detection was successful. In this study, we use the Braaten-Cohen-Yang cloud detector~\cite{Braaten2015} available from Sentinel Hub. In the example in Fig.~\ref{figCD}, a cloudy image is to the left and to the right is the same image with the Braaten-Cohen-Yang cloud detector applied.
Note that thin cirrus clouds and cloud shadows are not detected with this algorithm, but they would distort the data and this satellite image would not be suitable for analysing the water in the image.

\section{Results}
Getting a good match between the satellite images and in situ data is a common problem. For some uses it does not matter if there is a few days, or even weeks, difference between the satellite and in situ data. But the conditions in water can change fast, so a close match between the data is important.
We got data from NIVA at 55 different dates, and we only found eight suitable satellite images (not obscured by clouds) to match with the NIVA data. Sentinel-2 covers Mjøsa every 1-2 days, and only two of the eight dates match exactly. The rest had a one day difference between the satellite and NIVA data.

It should also be noted that the conditions of the water have great variations within a small area. If several samples are taken to look for concentration of algae within the same area, there can be a significant difference between the samples. This is one of the reasons the overview satellite imagery can give of a body of water can be so useful, but it also makes it hard to establish a ground truth reference for the satellite imagery.

\subsection{Se2WaQ}
The results from Se2WaQ is visual, with different colours marking the severity of the water quality parameters like turbidity and chl-a. The NIVA data is in numerical values for this parameters gathered at the different stations. The following is an example of NIVA data from the 9th of July 2019, refer to Fig. \ref{figNivaSta} for the station location:
\begin{itemize}
  \item M071 - Turbidity: 7,25398 – Chl-a: 1,7 
  \item M072 - Turbidity: 7,39289 – Chl-a: 2,3
  \item M073 - Turbidity: 0,9617 – Chl-a: 3 
  \item M074 - Turbidity: 0,55414 – Chl-a: 2,4 
\end{itemize}

\begin{figure}[t]
    \centering
    \subfloat[Turbidity]{\adjincludegraphics[width=.48\linewidth,trim={{.3\width} 0 {.2\width} 0},clip]{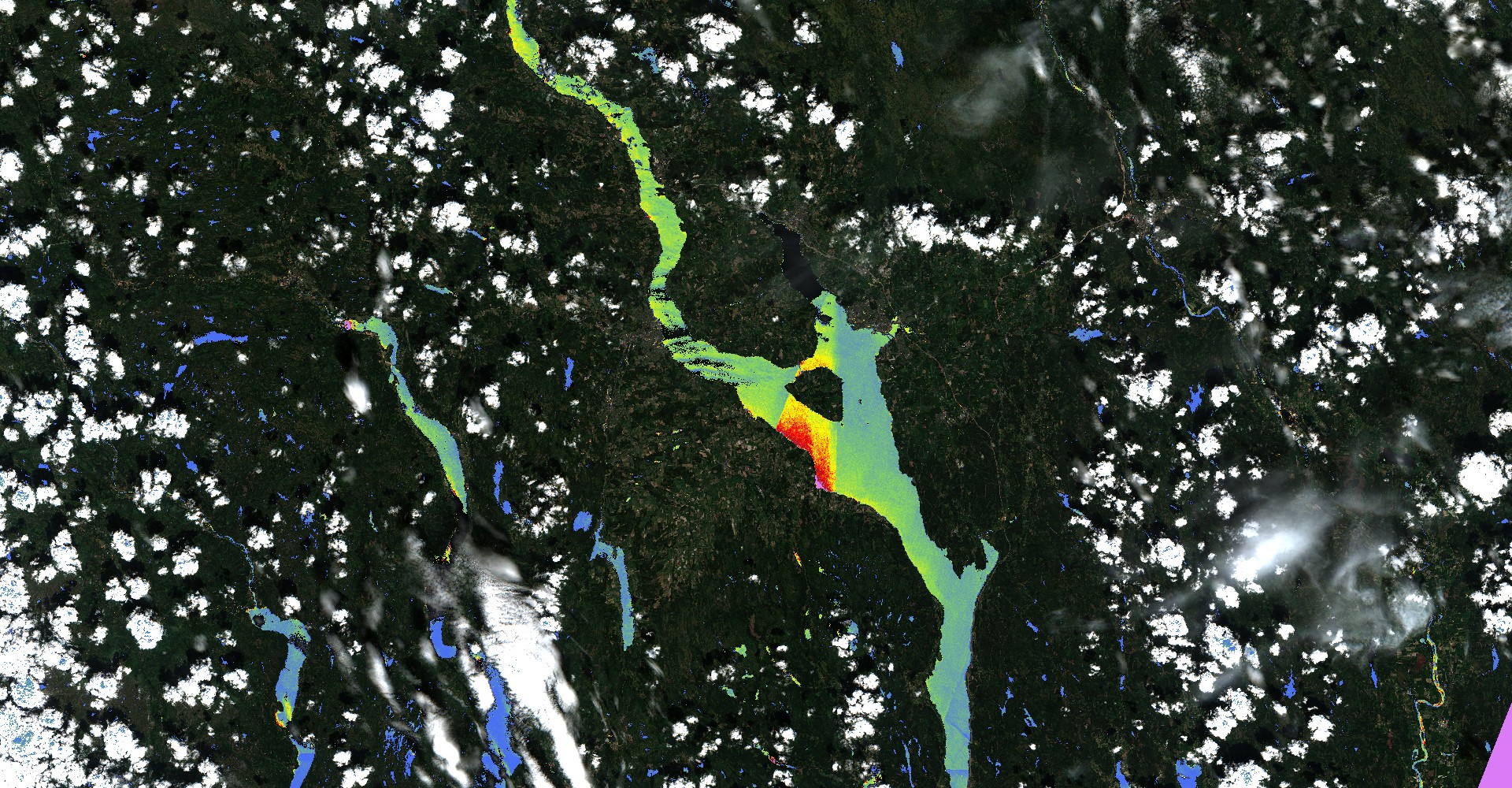}}\hfil
    \subfloat[Chl-a]{\adjincludegraphics[width=.48\linewidth,trim={{.3\width} 0 {.2\width} 0},clip]{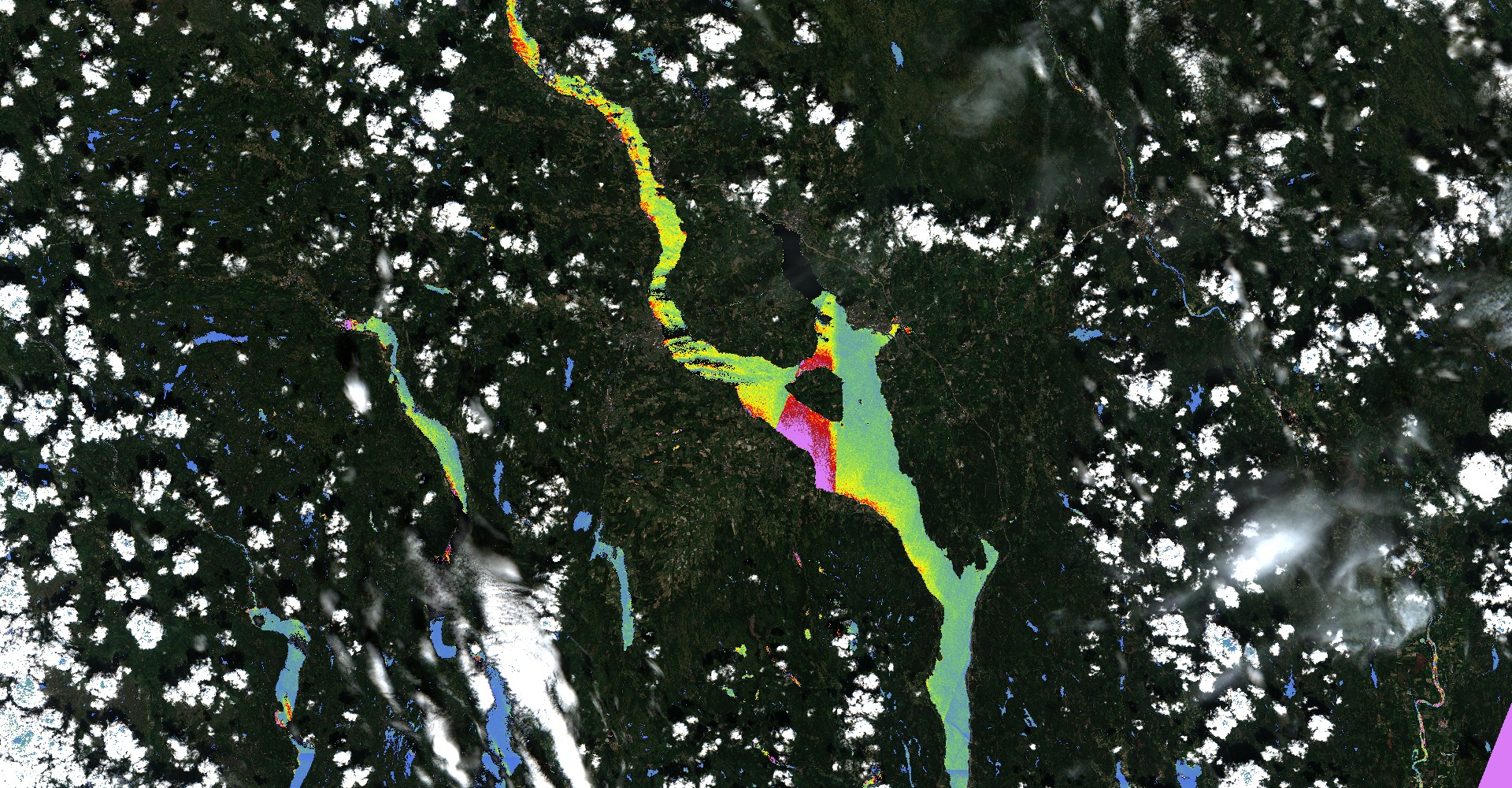}}
    \caption{Turbidity and Chl-a levels obtained using Se2WaQ script over Mjøsa on 10th June 2019.}
    \label{fig:se2waq}
\end{figure}

This was compared to satellite data from 10th of June, see Fig. \ref{fig:se2waq}. Like the NIVA data, Se2WaQ found more turbidity in the upper areas of Mjøsa near the M071 and M071 stations. But the difference between those stations and values from the M073 and M074 stations were not as strong in the Se2WaQ results. There was little to no correlation between the chl-a values in the 9th of July NIVA data and the 10th of July results from Se2WaQ. 

These results are typical of the comparisons between the NIVA data and the Se2WaQ results. Sometimes there is a fairly close correlation, and sometimes no correlation at all. There is not much data to compare with, but what correlation there is between the data is so weak that it might be coincidental.
There is also a chance that the weak correlation is a result of the NIVA data and the satellite data being taken at different times, but Se2WaQ showed no improvement on the two times when NIVA and satellite data was from the same day.

Se2WaQ is done directly in EO Browser and uses the Sen2Cor, which we now know is not the best atmospheric correction for analysing water quality. With this knowledge, and results, it seems fair to conclude that Se2WaQ in EO Browser is probably not suitable for estimating water quality in Mjøsa.

\subsection{APA}
\begin{figure}[ht]
    \centering
    \includegraphics[width=\linewidth]{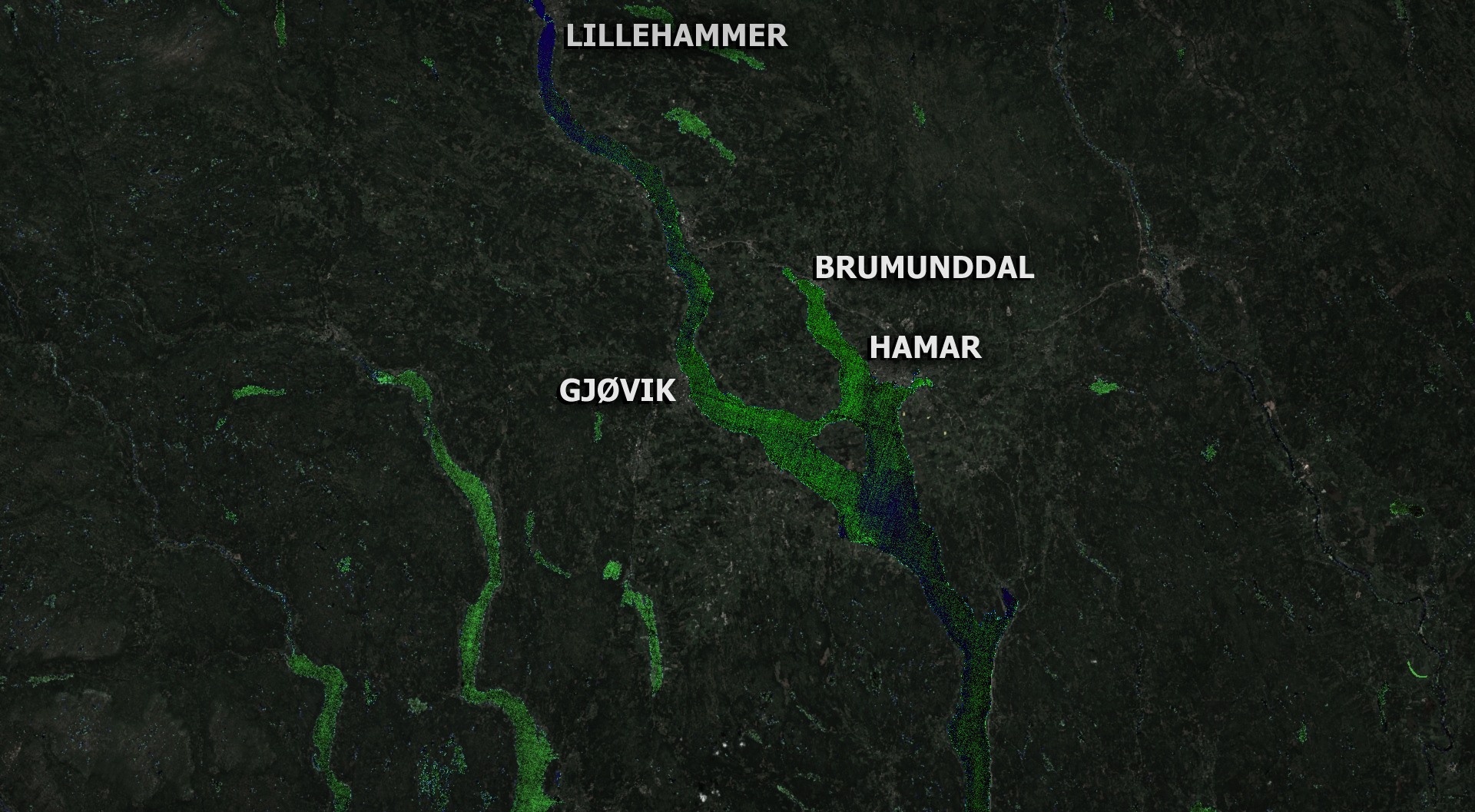}
    \caption{APA script result over Mjøsa on 16th July 2021.} 
    \label{fig:apa}
\end{figure}

With the APA script in EO Browser we tested the scripts on dates of reported algal blooms in Mjøsa. One such bloom happened in July 2021, and the script showed a clear increase in algae activity leading up to the bloom, see Fig. \ref{fig:apa}. The algae bloom was reported near Ringsaker, and was particularly severe by Brumunddal. But the APA script seemed to show the same high algae activity for almost all of Mjøsa. This is typical for our experience with the APA script. There is some correlation between the results from the scripts and dates of reported algae blooms, but the script does not show higher activity in the specific areas of Mjøsa where the algal blooms are reported.

The APA script also only gives very generic results, since it looks for algae and water plants in general. It has no information about what type of algae or plant is being detected. The APA scripts also detects high activity at times when there are no reports of algal blooms. This may be false positives, or that something other than algae is detected. It also occasionally detects high algal or water plant activity in May, when the ice on Mjøsa has just melted and it is very unlikely that there would be any such activity yet.

APA shows a little promise for monitoring algal blooms in Mjøsa, but is unreliable with its occasional detection of unlikely high algal or water plant activity in May. When it does work the results are imprecise and generic, and APA is probably not a reliable tool for studying algal activity in Mjøsa.

\section{Conclusion}
In this preliminary study we have gathered information about the usage of satellite imagery for analysing water quality to get an overview of the state of the art of the field. Lake Mjøsa was used as a case study and we compared our findings with in situ measurements provided by NIVA. We attempted to analyse water quality directly within EO Browser using the Se2WaQ and APA scripts. Neither of them performed particularly well. The Sen2Cor atmospheric corrected data available in EO Browser is not very suitable for analysing water quality and it is unlikely that Se2WaQ and APA, or any algorithms used directly in EO Browser, can be used to study the water quality of Mjøsa with a acceptable level of accuracy.

Quick analysis of water quality with tools like EO Browser or Google Earth Engine is probably still something for the future. The main problem is the need for specialized atmospheric correction when working with different types of water, but it is possible that data with types of different atmospheric correction will be made available directly within these already powerful browser applications as they are under continuous development.
For now, the best approach when wanting to analyse water quality is to download the full data, do atmospheric correction suitable to the type of water studied and analyse the data with tools like The European Space Agency's SNAP \cite{SNAP}.

When state of the art approach is used it should be possible to attain results that are useful. More work still needs to be done to increase the accuracy when studying water quality with satellite imagery, but progress is being made.

Mjøsa is changing and needs to be closely monitored. Satellite imagery can give a great overview of Mjøsa and complement in situ measurements. This kind of overview is important because water conditions may have great variations within a small area and a sample does not necessarily represent the condition of the area. In situ measurements of water quality are also time consuming and labor intensive, and satellite imagery can provide information about the water quality on days when in situ measurements are not done.

It should also be noted that few lakes in Norway, if any, have been as closely monitored as Mjøsa has been since the start of Mjøsaksjonen. This means there is a lot of data on Mjøsa that the satellite data can be compared with, and as such it is a suitable starting point for monitoring Norwegian lakes. But many lakes in Norway are in hard-to-reach locations and the water quality is only tested every 2-3 years. These lakes would also greatly benefit from the increased monitoring of water quality that satellite imagery can provide.

\section{Acknowledgment}
Authors would like to thank Therese Harvey from NIVA Denmark for the invaluable input and discussion, and also for providing field measurement data collected by NIVA used in this study.

\bibliographystyle{splncs04}
\bibliography{forArxiv.bib}

\newpage
\pagestyle{empty}
\noindent\makebox[\linewidth]{\rule{\linewidth}{0.4pt}}
\section*{Reviews received from NIKT 2021 (Rejection)}
\noindent\makebox[\linewidth]{\rule{\linewidth}{0.4pt}}
~\\
\noindent The manuscript was submitted to the 33rd Norwegian ICT Conference for Research and Education - \href{https://www.ntnu.edu/nikt2021}{\textcolor{blue}{NIKT 2021}} (NIK track). The following texts are the reviews, verbatim and only reformatted for ease of reading from an email format. 

\subsection*{Review 1}
\begin{tabular}{llll}
    Relevance for NIK && SCORE: & 2 (within scope)\\
    Originality && SCORE: & 2 (high)\\
    Readability && SCORE: & 1 (good)\\
    Technical quality && SCORE: & 1 (good)\\
    \textbf{Overall evaluation} && SCORE: & 2 (accept)\\
\end{tabular}
\\~\\
\noindent \textbf{Text:}

\noindent This paper describes various ways of analysing satellite images or lake Mjøsa to identify markers of water quality. The paper is clear and well presented. 
\\~\\
\noindent There should be a clarification that this is about image analysis early on in the paper or even in the title, as it is, I was unsure how this was relevant some way into the paper. This was however resolved later.
\\~\\
\noindent It is interesting to see that a field with so much potential has this many unexplored topics. Looking forward to see this project progress.

\subsection*{Review 2}
\begin{tabular}{llll}
    Relevance for NIK && SCORE: & 2 (within scope)\\
    Originality && SCORE: & 1 (medium)\\
    Readability && SCORE: & 1 (good)\\
    Technical quality && SCORE: & -2 (very poor)\\
    \textbf{Overall evaluation} && SCORE: & -1 (weak reject)\\
\end{tabular}
\\~\\
\noindent \textbf{Text:}

\noindent This paper describes using existing software and algorithms to quantify water quality in Mjøsa.
\\~\\
\noindent The paper provides a good background of water quality measurements, and the attributes measured.
\\~\\
\noindent Little related work in detecting algal blooms from satellite images is described, and the methodology is not compared to the state of the art.
\\~\\
\noindent The novelty and contribution of the paper is not described clearly.
\\~\\
\noindent No statistical evaluation has been performed on the results, so no clear conclusion is given.

\subsection*{Review 3}
\begin{tabular}{llll}
    Relevance for NIK && SCORE: & -1 (out of scope)\\
    Originality && SCORE: & 0 (low)\\
    Readability && SCORE: & 2 (excellent)\\
    Technical quality && SCORE: & 1 (good)\\
    \textbf{Overall evaluation} && SCORE: & -2 (reject)\\
\end{tabular}
\\~\\
\noindent \textbf{Text:}

\noindent This is a well-written paper on the possibilities of usingsatellite images to measure different kinds of pollution, and the algal and Cyaanobacterial blooming effects in lakes.
\\~\\
\noindent The practical qualitative isues are very well described. In fact they are so well described that the paper is very engaging.
\\~\\
\noindent But there is no statistical (or deterministic) framework described for describing potential solutions. Instead, one particular solution based on simple scripting is used, possibly becsue thsi is common in the geo-spatial imaging literature.
\\~\\
\noindent There is no algorithm mentioned. There is no signal processing described. Potential problems are mentioned, but without explaiining why they cannot be overcome. For example, stitching of images is described as aproblem. But it seems to this reviewer that once we know that stitching is present in an individual image, we can design and run algorithms.
\\~\\
\noindent There is no emphasis on designing algorithms. Instead, the paper merely reports what results from a standard scripting process.
\\~\\
\noindent Otherwise, the paper is very informative on the basic problem, using easy to understand language.

\end{document}